\documentclass[twocolumn,aps,pra,amsmath,amssymb]{revtex4-1}
\usepackage{physics}
\usepackage{graphicx}
\usepackage{mwe}
\usepackage{bm}
\usepackage{epstopdf}
\usepackage{subfig}
\usepackage{epsfig}
\usepackage{lipsum}
\usepackage{epstopdf}
\usepackage{epsfig}
\usepackage{dsfont}
\usepackage{amssymb}

\usepackage{braket}

\usepackage{leftidx}
\begin{document}

\title{Quantum Coherence Dispersion}

\author{Fernando Parisio}
\email[]{fernando.parisio@ufpe.br}
\affiliation{Departamento de
		F\'{\i}sica, Centro de Ci\^encias Exatas e da Natureza, Universidade Federal de Pernambuco, Recife, Pernambuco
		50670-901 Brazil}

\begin{abstract}
We investigate how quantum coherence can be distributed among the several off-diagonal elements of an arbitrary density matrix. An easily computable quantity that captures this variability notion  is proposed and it is argued that it presents marked features of complexity quantifiers. It turns out that this coherence dispersion ($\Delta_{\rm c}$) is maximized for intermediate values of an appropriate entropy (the relative entropy of coherence), a prevalent signature of complexity quantifiers across different fields, from evolutionary biology to linguistics and information science. The final part of the manuscript is dedicated to the connection between the proposed framework and non-equilibrium systems in the quantum regime.

\end{abstract}

\maketitle
\section{Introduction}

Several everyday experiences rely on coherence, such as the interference patterns on the surface of a pond \cite{ducks} or diffracted sunlight \cite{trees}. What challenges common sense and makes quantum coherence unique is that it allows for the coexistence of classically irreconcilable circumstances. This possibility of superposition is not only a cornerstone of quantum phenomena, but also a pervasive ingredient in new technologies. 
Therefore, it does not come as a surprise the enormous effort invested to characterize quantum coherence \cite{l1,girolami,yuan,adesso,rana,bu,xu,biswas,review} and relate it with other primeval resources \cite{review,streltsov,vedral,qi,hu,dong,min, ding} in the last decade. 
Studies on genuine multi-level coherence \cite{ringbauer, ernesto}, or how coherence distributes over separated subsystems \cite{byrnes}, e. g., providing a finer characterization, have been carried out more recently.
These works illustrate the complexity involved in a deeper description of coherence, beyond its bare quantification. 

Although notions related to complexity pervade all scientific disciplines, setting a theoretical framework with operational concepts is still an open problem in many fields. Such framework is well established in certain areas, such as information science, e. g., where quantitative definitions of  computational complexity exist. The formalization of complexity has also been explored in a multitude of fields including text analysis \cite{linguist,linguist2}, ecology \cite{env,acoustic}, and other complex systems \cite{hogg, complx1, complx2}. A common feature of complexity indicators is their dual relationship with disorder (entropy) \cite{hogg}. Typically, low complexity is linked with both maximal and minimal entropy, while maximal complexity occurs for intermediate disorder. 

In this context, it is natural to consider the dispersion of the coherences of a quantum state, in a given basis. We address this question and, in the first part of this work, propose an easily computable variability figure of merit, which captures essential features of  complexity quantifiers, being valid for composite systems with an arbitrary number of components. These features make coherence dispersion particularly suited to address complex, out-of-equilibrium systems comprised by a large number of parties. 

In the final part of this manuscript, we study a composite system comprised by a large number $n$ of $d$-level systems. We assume that each subsystem is described by coherent Gibbs states \cite{rudolph} after dephasing takes place and find that a curious effect emerges: coherence dispersion vanishes, except for a narrow temperature interval, which strongly depends on a characteristic energy scale and weakly depends on $n>>1$. Curiously, this finding is robust against decoherence.

\section{Coherence dispersion}
\label{S2}

Some well known quantities related to a density matrix can be seen under an alternative but elementary, statistical perspective. Consider an orthonormal basis $\{|i\rangle\}$ in a Hilbert space of finite dimension $D$ and a quantum state $\varrho$ in the associated Hilbert-Schmidt space $B({\cal H})$. 
Let us first address the populations, $\varrho_{ii}=\langle i|\varrho|i\rangle$. The constraint ${\rm Tr}(\varrho)=\sum_{i} \varrho_{ii}=1$ can be equivalently cast as follows: the mean population is fixed, $\overline{\varrho_{ii}}\equiv \overline{\rho}_{\rm p}=1/D$, where the over-score denotes average (${\rm Tr}(\varrho)=1\Leftrightarrow\overline{\rho}_{\rm p}=1/D$).
If, for whatever reason, one calculates an average, the natural next step is to determine the variance ($\sigma^2$). In the present case it reads $D \sigma^2_{\rm p}=\sum_{i}(\varrho_{ii}- \overline{\rho}_{\rm p})^2$. Interestingly, this is nothing but the squared predictability \cite{predic} (apart from a multiplicative factor): 
$$P^2=\sum^{D}_{i=1} \varrho_{ii}^2-\frac{1}{D}=D\sigma^2_{\rm p},$$
introduced in a quite distinct context, as a complementary quantity to visibility, in the field of quantum optics.

\begin{figure}
\begin{center}
  \includegraphics[height=3.8cm]{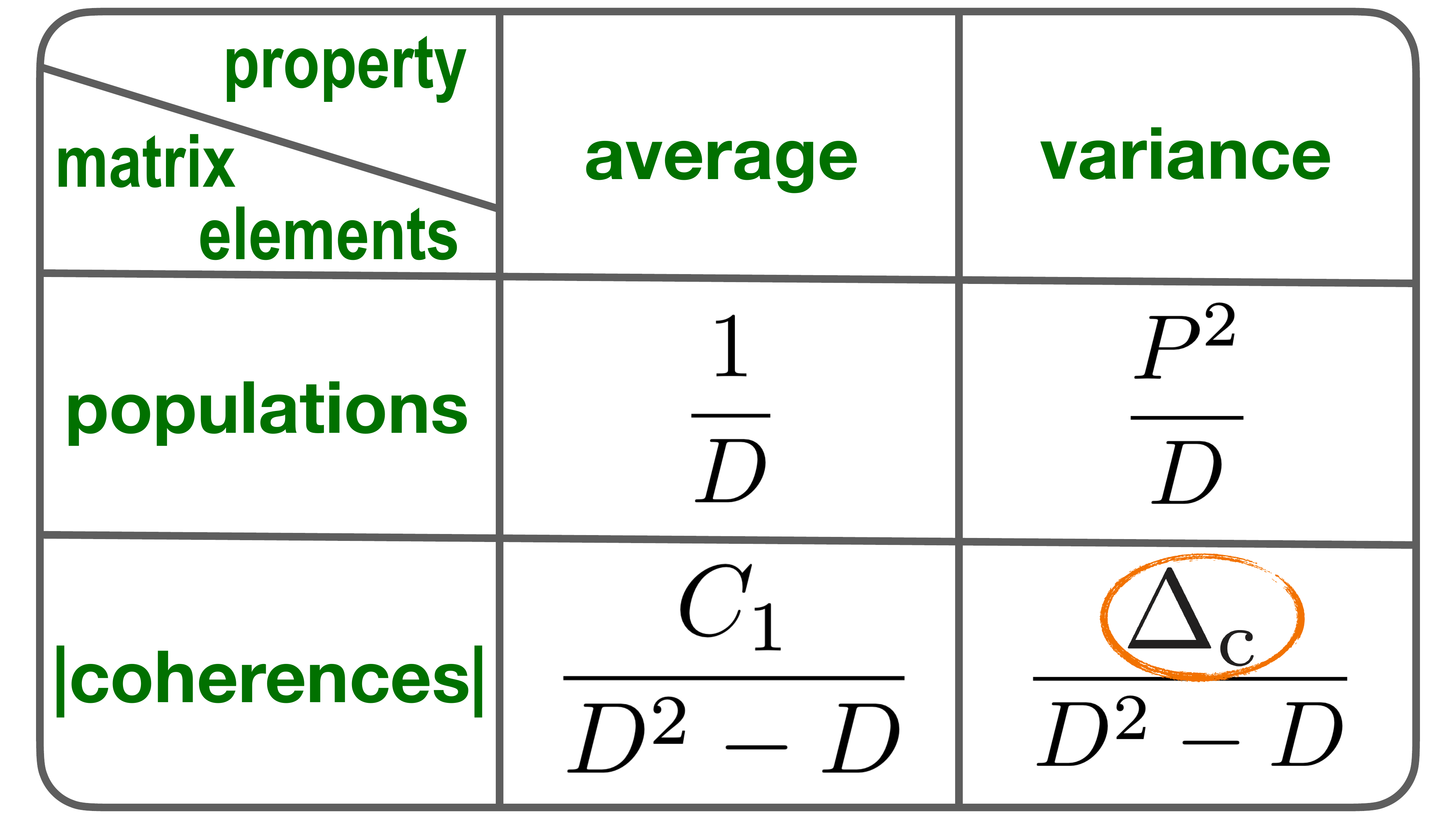}
  \caption{Summary of statistical properties of the entries of $\varrho$. Eq. (\ref{V}) fills the gap related to the variance of the absolute values of the coherences.}
  \label{fig1}
    \end{center}
\end{figure}

Now consider the absolute values of the $D^2-D$ coherences of $\varrho$. It is evident that the corresponding average $\overline{|\varrho_{ij}|}\equiv \overline{\rho}_{\rm c}$, $i\ne j$, is simply proportional to the $\ell_1$ norm of coherence ($C_1$):
\begin{equation*}
\overline{\rho}_{\rm c}=\frac{1}{D^2-D} \sum_{i\ne j}^D|\varrho_{ij}|=\frac{C_1(\varrho)}{D^2-D},
\end{equation*}
thus being a proper coherence monotone \cite{l1}. We are left with the variance of the numbers $|\varrho_{ij}|$ $(i\ne j)$, 
\begin{equation*}
\sigma^2_{\rm c}=\frac{1}{D^2-D}\sum_{i\ne j}^D(|\varrho_{ij}|-\overline{\rho}_{\rm c})^2\equiv \frac{{\Delta_{\rm c}}(\varrho)}{D^2-D},
\end{equation*}
 where we call ${\Delta_{\rm c}}$ the coherence dispersion.
One easily obtains:
\begin{equation}
\label{V}
{\Delta_{\rm c}}(\varrho)=C_2(\varrho)-\frac{\left(C_1(\varrho)\right)^2}{D^2-D},
\end{equation}
with $C_2(\varrho) = \sum_{i\ne j}^D|\varrho_{ij}|^2$ being the so-called $\ell_2$ ``norm'' of coherence. Eq. (\ref{V}) fills the otherwise blank entry in Fig. \ref{fig1}.
Note that ${\Delta_{\rm c}}=0$ for any incoherent state but also for any maximally coherent state. Other properties follow directly from the general features of variances, for instance,  \cite{dahl}: 
\begin{equation}
\nonumber
{\Delta_{\rm c}}\left(\sum_k p_k \varrho_k \right)\le \sum_k p_k {\Delta_{\rm c}}(\varrho_k).
\end{equation}
For completeness, in appendix \ref{A1}, we provide a demonstration and also show that $0 \le {\Delta_{\rm c}} < 1$, for any dimension $D$, with the upper bound being approached for $D$ arbitrarily large. So coherence dispersion is not an extensive quantity.

Note also that ${\Delta_{\rm c}}$ vanishes for the simplest systems:  those with $D=2$.
Coherence Dispersion displays other traits of a complexity indicator, due to its relation with the appropriate disorder figure of merit, namely,  the relative entropy of coherence \cite{l1}. It is associated with coherences only and given by 
\begin{equation*}
S_{\rm c}(\varrho)=S(\varrho_{\rm diag})-S(\varrho),
\end{equation*}
where $S$ in the von Neumann entropy and $\varrho_{\rm diag}$ is the incoherent state with the same populations as $\varrho$. Therefore, for an arbitrary pure state $\psi=|\psi\rangle \langle \psi |$ we have  $S_{\rm c}(\psi)=S(\psi_{\rm diag})$, since $S(\psi)=0$.

\begin{figure}
 \setbox1=\hbox{\includegraphics[height=2.6cm]{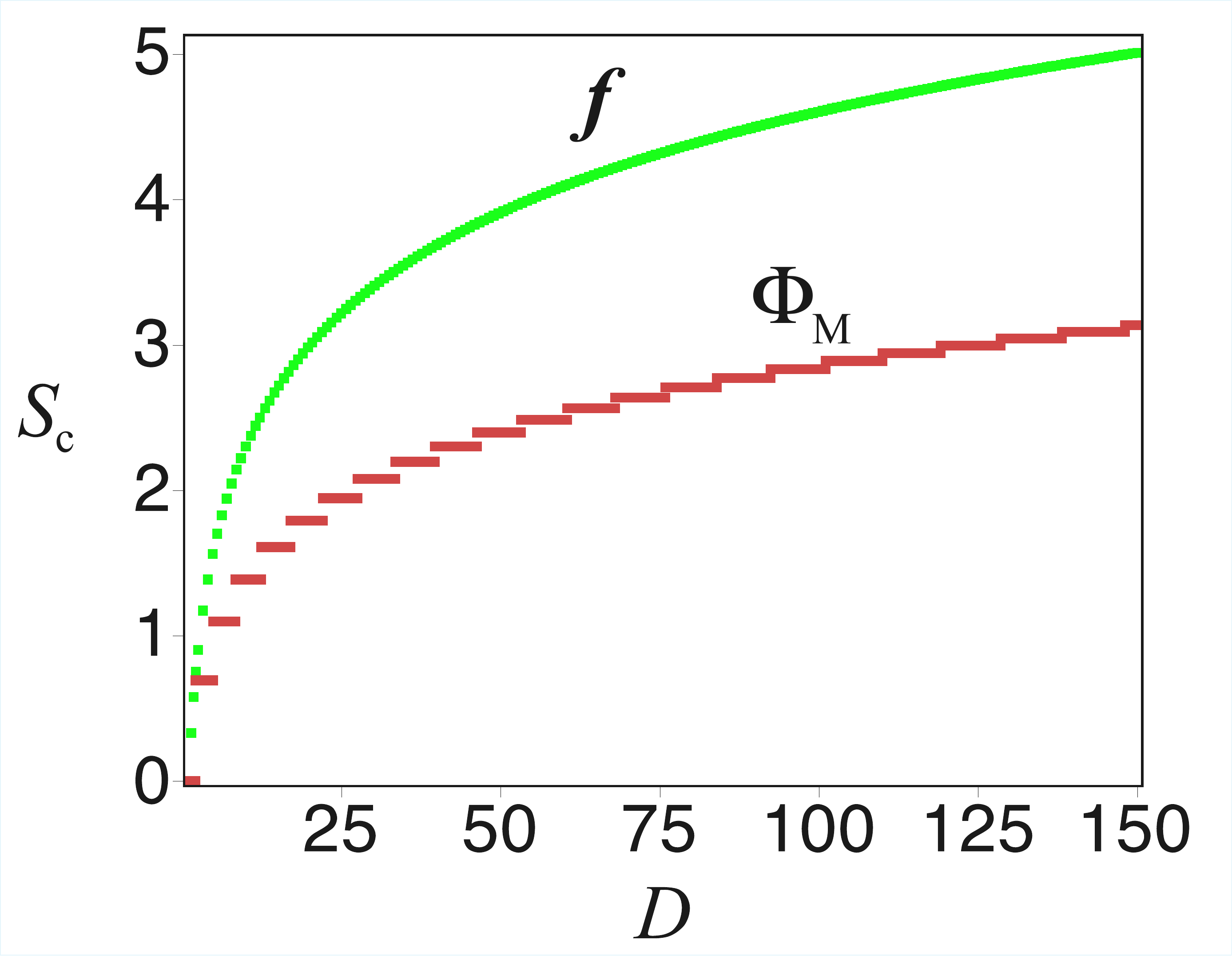}}
  \includegraphics[height=4.9cm]{fig2.pdf}\llap{\makebox[\wd1][l]{\raisebox{0.9cm}{\includegraphics[height=1.65cm]{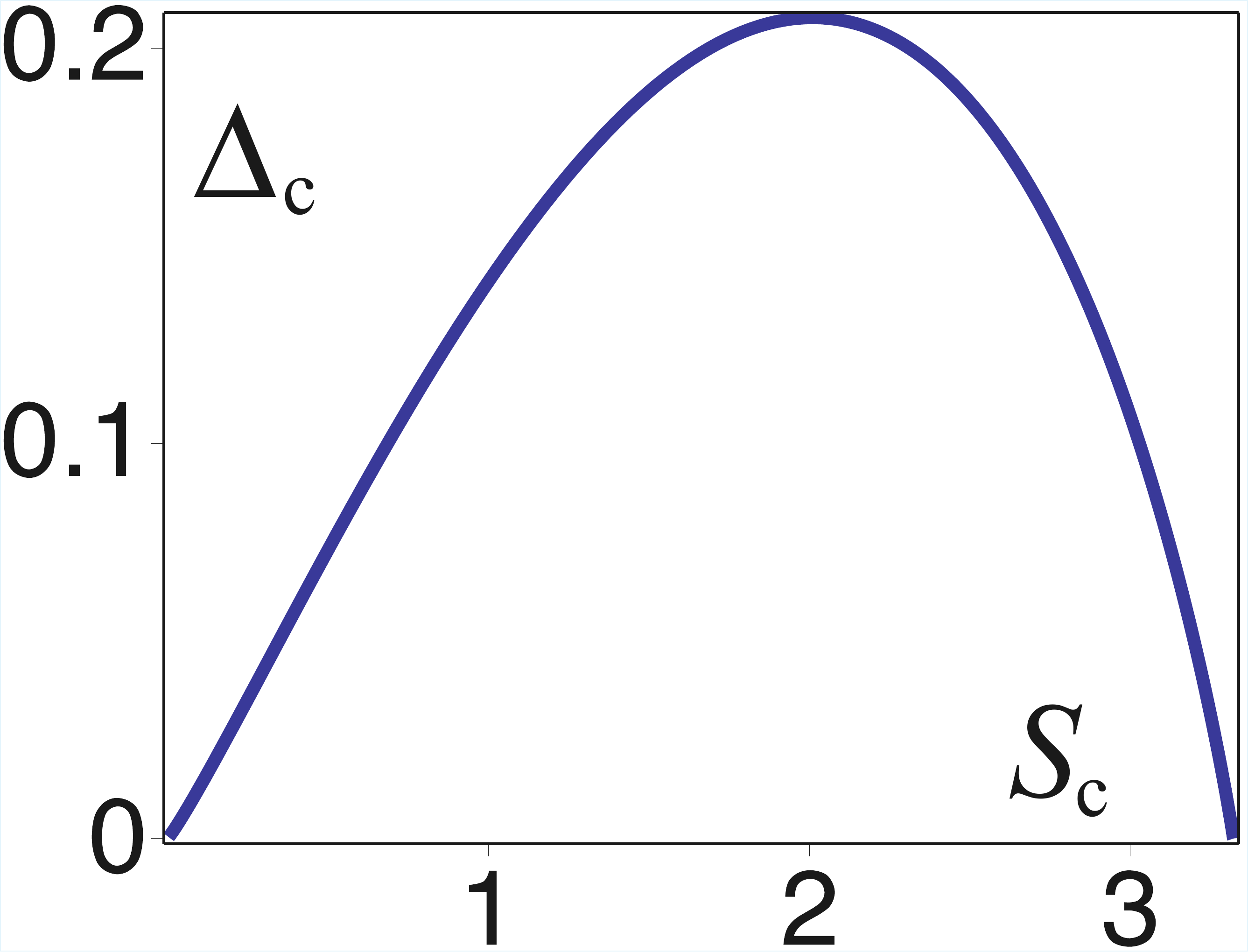}}}}
  \caption{$S_{\rm c}$ against the dimension $D$, for the states $f$ and $\Phi_{\rm M}$. Inset: ${\Delta_{\rm c}}$ versus $S_{\rm c}$ for the state $|\psi_x\rangle \langle \psi_x|$ with $D=10$. All plotted quantities are dimensionless.}
  \label{fig2}
\end{figure}

It is easy to see that, within pure states, one can reach minimal and maximal values of $S_{\rm c}$. If the ket is one of the elements of the selected basis, say $|1\rangle$, then $S_{\rm c}=0$; while, for $|f\rangle=(1/ \sqrt{D})\sum_i|i\rangle$ we get $S_{\rm c}=\log D$. It is immediate that in both cases we obtain ${\Delta_{\rm c}}=0$. Therefore, ${\Delta_{\rm c}}$ vanishes for maximal and minimal $S_{\rm c}$. This is an ubiquitous signature of complexity indicators in a variety of fields \cite{linguist,linguist2,env,acoustic,hogg,complx1,complx2}. In addition, complexity figures of merit attain a single maximum at a some intermediate entropy value. Due to convexity, it is clear that states which maximize ${\Delta_{\rm c}}$ must be pure ($|\Phi_{\rm M}\rangle$). Indeed, it is always possible to express an arbitrary state $\varrho$ in terms of a convex sum of pure states, $\varrho=\sum p_i |\varphi_i\rangle \langle \varphi_i|$. From convexity, ${\Delta_{\rm c}}(\varrho)\le \sum_i p_i {\Delta_{\rm c}}(\varphi_i)$ and ${\Delta_{\rm c}}(\varphi_i)\le {\Delta_{\rm c}}(\Phi_{\rm M})$, by hypothesis. Therefore, ${\Delta_{\rm c}}(\varrho)\le {\Delta_{\rm c}}(\Phi_{\rm M})$, for any non-pure $\varrho$. The states $|\Phi_{\rm M}\rangle$ can be determined via a constrained optimization, with Lagrange multipliers. In appendix \ref{A2} we show that these maximum-dispersion states assume the form: 
\begin{equation}
\label{optimal}
|\Phi_{\rm M}\rangle=\frac{1}{\sqrt{r(D)}}\sum_{i=1}^{r(D)}|i\rangle,
\end{equation}
where $r(D)<D$ is an involved integer function of the dimension $D$ [e.g., $r(3)=r(4)=2$, $r(10)=4$, and $r(100)=17$], which comes from the cubic equation
\begin{equation*}
s^3-s^2=\frac{D^2-D}{2},
\end{equation*}
presenting a single real root $\mathfrak{s}$. The extremal rank is then
$r(D)= \lfloor \mathfrak{s} \rfloor +(1+{ \rm sign}(X))/2$,
with
\begin{equation*}
 X \equiv \frac{1}{\lceil  \mathfrak{s} \rceil\lfloor  \mathfrak{s} \rfloor}-\left(\frac{\lceil  \mathfrak{s} \rceil+\lfloor  \mathfrak{s} \rfloor-2}{D^2-D} \right),
\end{equation*}
where one writes the closest smaller integer as $\lfloor  \mathfrak{s} \rfloor$ and the closest larger integer as $\lceil  \mathfrak{s} \rceil$ (see appendix \ref{A2} for details).
State (\ref{optimal}) presents intermediate entropy of coherence, $S_{\rm c}(\Phi_{\rm M})=\log r(D) $, as can be seen in Fig. \ref{fig2} in comparison with maximal-entropy states $f$, with $S_{\rm c}(f)=\log D$.

This behavior is not limited to states with maximum dispersion. Consider, for instance, the family of normalized states 
\begin{equation*}
|\psi_x\rangle=a(x)|1\rangle+b(x)\sum_{i>1}|i\rangle, 
\end{equation*}
with
$a(x)=1-x, \;\; b(x)=\sqrt{\frac{x(2-x)}{D-1}}$,
which interpolates between $|1\rangle$ ($x=0$) and $|f\rangle$ ($x=1$), as $x$ goes from 0 to 1. The state $\psi_x$ never coincides with $\Phi_{\rm M}$, but as we vary $x$, some constrained maximum-dispersion state is attained, as shown in the inset of Fig. \ref{fig2}. The curve shows the archetypical dependence of complexity-entropy diagrams \cite{linguist,linguist2,env,acoustic,hogg,complx1,complx2}.

Since complexity is often an emergent property, it would be interesting to express the coherence dispersion of a composite system, comprising a large number of subsystems. We now show that ${\Delta_{\rm c}}$ is also computable in this sense:  one can easily determine ${\Delta_{\rm c}}(\rho^{\otimes n})$, for any number of copies $n$, from quantities easily obtainable from a single copy.  

For an arbitrary quantum mechanical figure of merit $F: \cal{B(H)}\to \mathds{R}$ and an arbitrary state $\rho$, the knowledge of $F(\rho)$ does not ensure the knowledge of $F(\rho^{\otimes n})$. A benign exception is that of additive functions, for which we simply have $F(\rho^{\otimes n})=nF(\rho)$, as is the case for the logarithmic negativity \cite{vidal,plenio} (for entanglement) and rugosity \cite{texture} (for quantum-state texture). More generally, even if additivity does not hold, there are quantifiers $F$ for which the knowledge of $F(\rho)$ does ensure the knowledge of $F(\rho^{\otimes n})$, additivity being a particular case. Quantifiers $F$ that satisfy this requirement have been termed 1-scalable (1S) functions \cite{parisio}. It turns out that ${\Delta_{\rm c}}$ can be completely expressed in terms of 1-scalable quantities, namely, $P^2$,  $C_1$, and the purity $\Pi\equiv {\rm Tr}(\rho^2)$. This is done by employing the complementarity relation  \cite{compl}, 
\begin{equation*}
C_2+P^2-\Pi+\frac{1}{D}=0
\end{equation*}
to eliminate $C_2$, which is not scalable \cite{lucas}.  
For $\varrho=\rho^{\otimes n}$, the purity can be written as
\begin{eqnarray*}
\Pi(\varrho)&=&\Pi(\rho^{\otimes n})= \sum^{d}_{i_1 \cdots i_n=1 \atop j_1 \cdots j_n=1}|\rho_{i_1 j_1}\cdots \rho_{i_n j_n}|^2\\
&=&\sum^{d}_{i_1 \cdots i_n=1 \atop j_1 \cdots j_n=1}|\rho_{i_1 j_1}|^2\cdots |\rho_{i_n j_n}|^2\\
&=&\prod_{k=1}^n\left( \sum_{i_k,j_k}^d |\rho_{i_kj_k}|^2\right)= \left( \sum_{i,j}^d |\rho_{ij}|^2\right)^n=[\Pi(\rho)]^n.
\end{eqnarray*}

For the predictability we get
\begin{eqnarray*}
P^2(\varrho)&=&P^2(\rho^{\otimes n})=\sum^{D}_{i=1}|\varrho_{ii}|^2-\frac{1}{D}\\
&=&\sum^{d}_{i_1 \cdots i_n=1}|\rho_{i_1 i_1}|^2\cdots |\rho_{i_n i_n}|^2-\frac{1}{d^n}\\
&= &\left( \sum_{i=1}^d |\rho_{ii}|^2\right)^n-\frac{1}{d^n}=\left(P^2(\rho)+\frac{1}{d}\right)^n-\frac{1}{d^n}.
\end{eqnarray*}
In summary, for $\varrho=\rho^{\otimes n}$ we have:
$C_1(\rho^{\otimes n})=(1+C_1(\rho))^n-1$, $P^2(\rho^{\otimes n})=(P(\rho)^2+1/d)^n-1/d^n$,
and $\Pi(\rho^{\otimes n})=[ \Pi(\rho) ]^n$, where $\rho$ is represented by a $d \times d$ matrix, so $D=d^n$. 
The expression for the multi-copy $\ell_1$ norm of coherence is derived in \cite{lucas}. 

Thus, the $n$-copy coherence dispersion reads:
\begin{equation}
 {\Delta_{\rm c}}(\rho^{\otimes n})= \Pi^n -\left(P^2+\frac{1}{d} \right)^n-\frac{[(1+C_1)^{n}-1]^2}{ d^{2n}-d^n},
 \label{Vn}
\end{equation}
where all elementary quantities refer to a single copy: $\Pi\equiv \Pi(\rho)$, $P\equiv P(\rho)$, and $C_1 \equiv C_1(\rho)$. The previous result is indeed more general. Consider the  state $\varrho=\rho^{(1)}\otimes\rho^{(2)}\otimes\cdots \otimes \rho^{(n)}$, satisfying $\rho^{(k)}_{ij}=e^{\phi_{k\ell}}\rho^{(\ell)}_{ij}$, where $\phi_{k\ell}$ are arbitrary phases and $k,\ell \in \{1, \cdots, n\}$. Then, since $\Pi$, $P$, and $C_1$ only depend on the absolute values of the matrix entries, it follows that 
$V(\rho^{(1)}\otimes\rho^{(2)}\otimes\cdots \otimes \rho^{(n)})=V({\rho^{(1)}}^{\otimes n})$. So, if the states differ by local phases, Eq. (\ref{Vn}) can still be used. The same conclusion holds for states that differ by arbitrary permutations of the basis elements.
Finally, note that coherence dispersion can be super-activated. A simple instance is as follows: any two-qubit state $\sigma$ has ${\Delta_{\rm c}}(\sigma)=0$, but we usually have ${\Delta_{\rm c}}(\sigma^{\otimes n})\ne 0$, for $n>1$.

\section{Coherence dispersion and pure quantum athermality}
\label{S3}

As an illustration of how coherence dispersion behaves in a case of potential interest, we investigate how $\Delta_{\rm c}$ relates to thermodynamical concepts. Consider a system in thermal equilibrium with a reservoir at absolute temperature $T$. It is described by the state
\begin{equation*}
\rho_G(T)=\frac{1}{\cal Z}\sum_i e^{- E_i/k_B T}|i\rangle \langle i|,
\end{equation*}
where ${\cal Z}$ is the canonical partition function, $k_B$ is the Boltzmann constant, and $\{|i\rangle\}$ is the energy basis, which also plays a central role in the resource theory of athermality \cite{atherm}. Since the Gibbs state has no coherences, we simply obtain ${\Delta_{\rm c}}(\rho_G)=0$. Also, incoherent states with populations that deviate from the Maxwel-Boltzmann (MB) distribution have zero dispersion. We say that these states display {\it classical athermality}.

Of course, ${\Delta_{\rm c}}$ can only appear in states with non-zero coherences, thus, being out of equilibrium in a quantum sense (athermality stemming from coherences). With this in mind we introduce the following concept. A state presents pure quantum athermality if, in the energy eigenbasis, its populations coincide with that of thermal states, i. e., $\rho_{ii}=\exp(- E_i/k_B T)/{\cal Z}$, and for $i\ne j$ we have $\rho_{ij}\ne 0$, for at least one pair $i,j$.

One class that is particularly interesting is that of coherent Gibbs states \cite{rudolph}, given by 
\begin{equation*}
|\Psi_{ G}\rangle=\frac{1}{\sqrt{\cal Z}}\sum_i e^{- E_i/2k_B T}|i\rangle,
\end{equation*}
which, under full decoherence, become $\rho_G$. 
We may use these states to describe open systems, which are in contact with a thermal bath, but not in equilibrium with it. 
Of course, a pure state is a contrived assumption for an open system. In this context, it is more natural to consider partially coherent Gibbs states: 
\begin{equation}
\label{G}
G=\lambda |\Psi_G\rangle \langle \Psi_G|+(1-\lambda)\rho_G.
\end{equation}
Note that, since the populations obey the MB distribution, these states present pure quantum athermality.
In this case, the single-copy coherence dispersion can be compactly expressed as (see appendix \ref{A3}):
\begin{equation*}
{\Delta_{\rm c}}\left(G(T)\right)=\lambda^2\left(1-\Pi_{\rm eq}(T)\right)-\frac{\lambda^2}{d^2-d}\left( 1-1/\Pi_{\rm eq}(2T) \right)^2,
\end{equation*}
where $\Pi_{\rm eq}(T)$ is the purity of the thermal-equilibrium Gibbs state.
The dispersion ${\Delta_{\rm c}}(G(T))$ of the non-equilibrium system in contact with a reservoir at temperature $T$, depends on the purity associated with the corresponding equilibrated system at two different temperatures ($T$ and $2T$), 
a signature of athermality. Notice that $\lambda^2$ is a global multiplicative factor, in particular, not affecting the location of extremal points, whenever they exist.

In appendix \ref{A3} we also show that the corresponding $n$-copy coherence dispersion is given by:
\begin{eqnarray}
\label{VGZMC}
\nonumber
{\Delta_{\rm c}} ( G^{\otimes n}(&T&))= \left(\lambda^2+(1-\lambda^2)\Pi_{\rm eq}(T)\right)^{n}-\Pi_{\rm eq}(T)^n\\
&-&\frac{1}{d^{2n}-d^n}\left(\left(1-\lambda+\lambda /\Pi_{\rm eq}(2T)\right)^n-1\right)^2.
\end{eqnarray}
Therefore we reached an expression for coherence dispersion of an out-of-equilibrium system comprised by an arbitrary number of elementary entities. The system is in contact with an environment at absolute temperature $T$, and, importantly, may bear some remanent level of coherence, $\lambda$. 

\subsection{Systems with a single energy scale}

\begin{figure}
  \includegraphics[height=4.9cm]{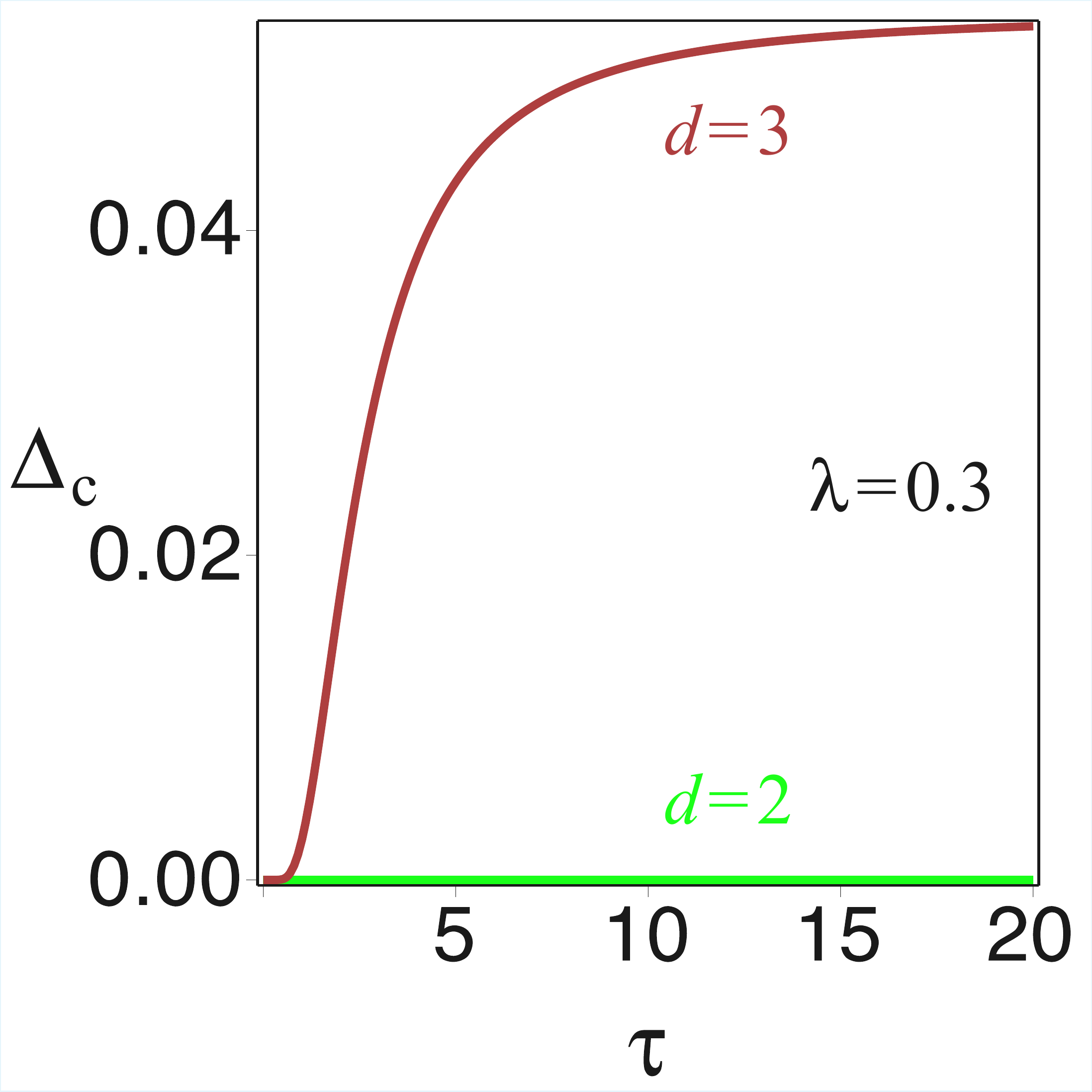}
  \caption{Single-copy $\Delta_{\rm c}$ against the dimensionless temperature $\tau$. For $d=2$ the coherence dispersion vanishes, while for $d=3$ it is a monotonically increasing function, saturating at a finite value.}
  \label{fig3}
\end{figure}

As an illustration, hereafter we will focus on systems with a single characteristic energy scale ($\epsilon$) and presenting pure quantum athermality. In appendix \ref{A3} we derive an expression for ${\Delta_{\rm c}}$ for $n$ such systems, which can exchange energy with the bath in packets of energy $\epsilon$, i. e., $E_j=j\epsilon$, $j=1\dots, d$, implying 
\begin{equation*}
\Pi_{\rm eq}(\tau) =\tanh( \frac{2}{\tau}) \coth(\frac{2d}{\tau}),
\end{equation*}
where $\tau$ is the dimensionless temperature given by 
\begin{equation}
\tau=\frac{4k_BT}{\epsilon}. 
\end{equation}
By plugging this into Eq. (\ref{VGZMC}) we find that for a single copy ($n=1$) and $d>2$, $\Delta_{\rm c}$ grows monotonically from $0$ at $\tau=0$ to a saturation value as $\tau \rightarrow \infty$, see Fig. \ref{fig3}.
\begin{figure}
  \includegraphics[height=4.9cm]{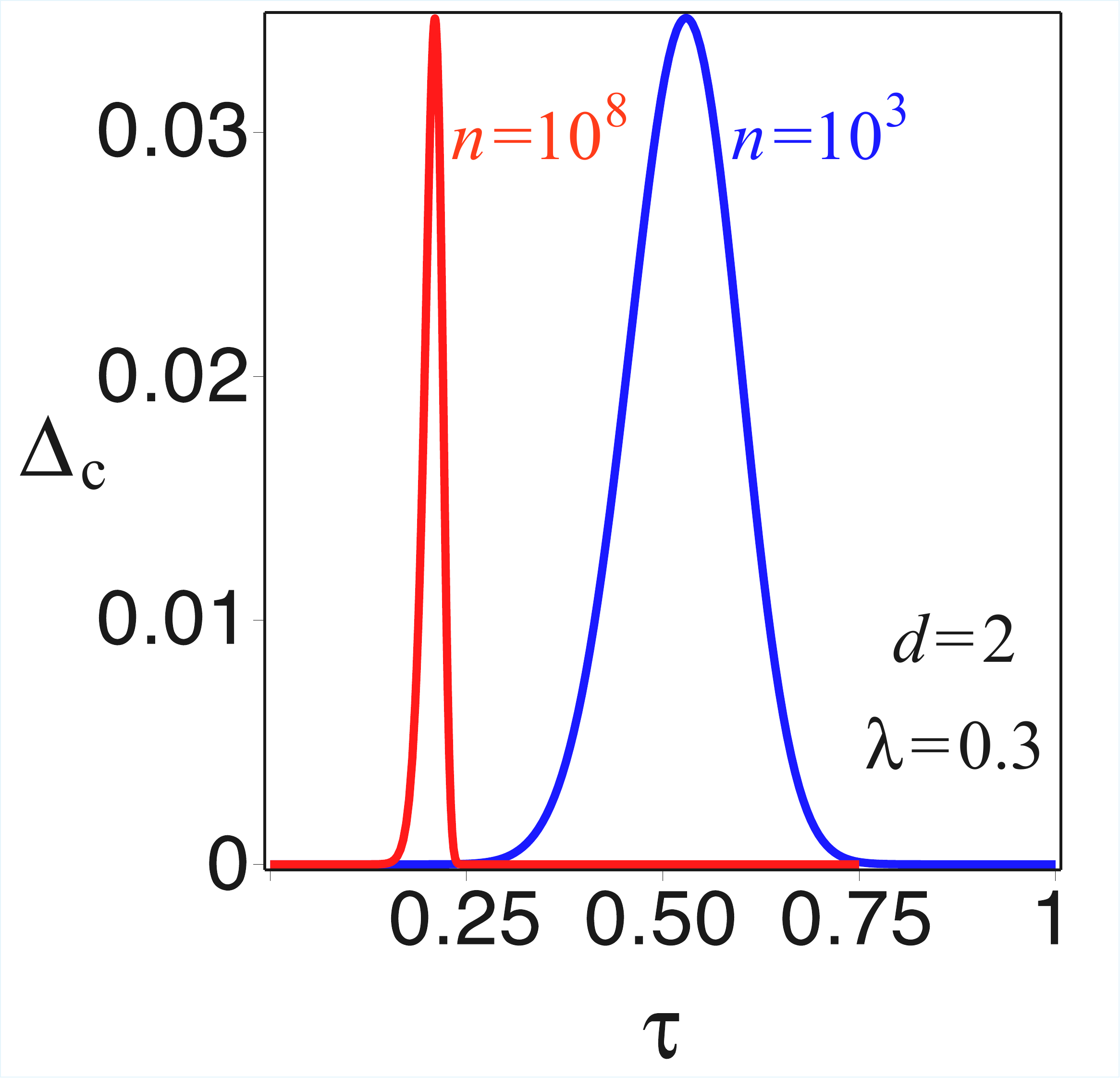}
  \caption{$\Delta_{\rm c}$ against the dimensionless temperature $\tau$, for $n=1000$ (right) and $n=10^8$ (left). The variation of five orders of magnitude in $n$ changes the temperature of maximal coherence dispersion from $\tau^*=0.53$ to $\tau^*=0.21$.}
  \label{fig4}
\end{figure}

\begin{figure}
  \includegraphics[height=4.9cm]{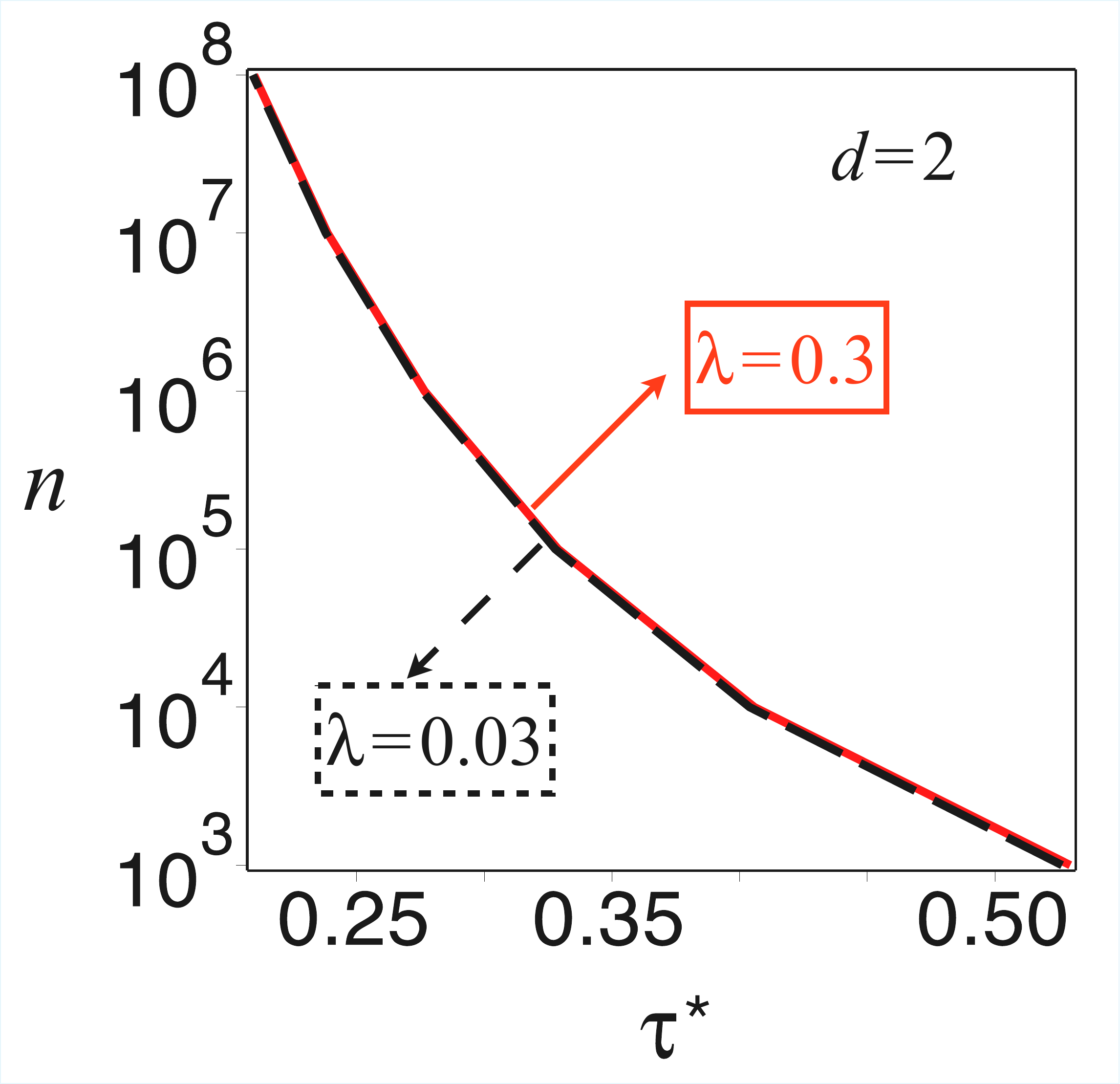}
  \caption{A logarithmic plot of $n$ versus $\tau^*$ for $d=2$ and $\lambda=0.3$ (continuous line) and $\lambda=0.03$ (dashed line). The two plots are very similar, illustrating the robustness of the values of $\tau^*$ against decoherence.}
  \label{fig5}
\end{figure}

Already for a modest number of copies ($n \gtrapprox10$), a distinctive property emerges: ${\Delta_{\rm c}}$ presents a single, sharp global maximum at a finite temperature $\tau^*$. In Fig. \ref{fig4} the behavior of $\Delta_{\rm c}$ as a function of $\tau$ is shown for $\lambda=0.3$ and $d=2$. The broader curve on the right corresponds to $n=1000$, while the sharper peak on the left corresponds to $n=10^8$. Coherence dispersion essentially vanishes for all temperatures, except in the vicinity of $\tau^*$, therefore sorting out a short temperature band. The position of the maximum is largely independent of $d$: had we set $d=30$ instead of $d=2$, the plot in Fig. \ref{fig4} would be visually indistinguishable with the difference in the values of $\tau^*$ appearing in the fifth decimal place. Interestingly, these results are also robust against decoherence, $\tau^*$ is weakly dependent of the coherence level $\lambda$ (of course the intensity of the peak does depend on $\lambda$). This feature can be observed in Fig. \ref{fig5}, where the values of $\tau^*$ are depicted for $10^3\le n \le 10^8$. The continuous line corresponds to $\lambda=0.3$ and the dashed line corresponds to $\lambda=0.03$. The largest difference between the two cases, in the investigated range, occurs for $n=1000$, with $\tau^*=0.5296$, for $\lambda=0.3$; and  $\tau^*=0.5263$, for $\lambda=0.03$.

To have a flavor of the magnitude of these emergent temperatures, let us use some typical figures. Consider a generic system with $n \sim 10^6$ elementary entities, which could, for instance, be electrons (between two bands) or spins (up or down). Let the energy gap be $\epsilon \approx 0.5 \; {\rm eV}$, compatible with several molecular and electronic processes. This gives
$T  \approx   \tau \times 1300\; \mbox{K}$.
For $n = 10^6$, a typical number in mesoscopic systems, we get the approximate dimensionless temperature of $\tau^* \approx 0.277$, or $ T \approx 87^{\circ}{\rm C}$. If one considers the width at half height, the corresponding temperature window is approximately $48^{\circ}{\rm C}<T<117 ^{\circ}{\rm C}$. Roughly, the energy scale we used corresponds to semiconductors with narrow band gaps, as, for instance, lead sulfide (PbS) with $\epsilon \approx 0.41$ eV. In this example we would have $10^6$ electrons (or holes) in a superposition state (regarding the two bands). This, of course, should correspond to a (possibly small) fraction of the total number of charge carriers. Of course, whether the maximization of $\Delta_{\rm c}$ for a narrow temperature interval presents some measurable effect is an open question that may be addressed in future work.

\section{Closing Remarks}
\label{S5}

Coherent superpositions are the essential structure behind quantum phenomena and its characterization should be multifaceted, going beyond its bare quantification. Several advances in the direction of a more comprehensive approach to understand quantum coherence have been reported \cite{ringbauer,ernesto,texture,machado} and the present work intends to be a contribution to this program.
 
The quantity proposed and explored here, arguably, corresponds to a ``missing'' concept in the elementary statistics of populations and coherences (Fig. \ref{fig1}). 
Its connection with the notion of complexity seems to be worth of further investigation.
In addition, the fact that coherence dispersion easily scales to an arbitrary number of copies, opens the possibility to explore its connection with quantum thermodynamics and complex quantum systems.
It is therefore expected that coherence dispersion may be fruitfully applied in other branches of quantum foundations and information. 

A more comprehensive analysis of the physical meaning, if any, of the emergent temperature windows reported in the previous section, although outside the scope of the present work, may prove worthwhile. 
Another potentially interesting direction for future research is to study the dynamical properties of coherence dispersion or, more generally, how it transforms under general channels (completely positive and trace preserving maps).
Finally, a non-trivial open problem that deserves attention is whether a formal resource-theoretical framework can be set for coherence dispersion.

\begin{acknowledgments}
The author thanks P. R. A. Campos and M. Copelli for their comments and suggestions on this work. This research received financial support from the Brazilian agencies Coordena\c{c}\~ao de Aperfei\c{c}oamento de Pessoal de N\'{\i}vel Superior (CAPES), Conselho Nacional de Desenvolvimento Cient\'{\i}fico  e Tecnol\'ogico (CNPq), Funda\c{c}\~ao de Amparo \`a Pesquisa do Estado de S\~ao Paulo (FAPESP - Grant 2021/06535-0), and Funda\c{c}\~ao de Amparo \`a Ci\^encia e Tecnologia do Estado de Pernambuco (FACEPE - Grant BPP-0037-1.05/24).
\end{acknowledgments}

\appendix
\widetext
\section{Convexity of ${\Delta_{\rm c}}$}
\label{A1}
For completeness, we give a proof of the convexity of 
\begin{equation}
\label{CD}
{\Delta_{\rm c}}=C_2-\frac{C_1^2}{D^2-D}.
\end{equation}
To simplify the notation, consider the following vectorization. Let ${\bf x}=(|\varrho_{12}|, |\varrho_{13}|, \dots, |\varrho_{1D}|, |\varrho_{23}|, |\varrho_{24}|, \dots, |\varrho_{(D-1)D}|) \in {\cal UC}_M \subset \mathds{R}^M$, with $M=(D^2-D)/2$ (the number of above-diagonal coherences). The set ${\cal UC}_M$ is the $M$-dimensional  unit cube ($0\le x_i\le1$). With this we simply have $C_1(\varrho)=C_1({\bf x})=2\sum_{j=1}^Mx_j$ and $C_2(\varrho)=C_2({\bf x})=2\sum_{j=1}^Mx_j^2$. Therefore,
\begin{equation}
\label{CD}
{\Delta_{\rm c}}({\bf x})=2\sum_{j=1}^Mx_j^2-\frac{4}{D^2-D}\left( \sum_{j=1}^Mx_j\right)^2.
\end{equation}
Note that the domain of definition of ${\Delta_{\rm c}}({\bf x})$ is convex since, $p{\bf x}+(1-p){\bf y} \in {\cal UC}_M$ for any pair ${\bf x}$, ${\bf y} \in {\cal UC}_M$. Of course ${\Delta_{\rm c}}$ is a twice-differentiable function of its argument and the corresponding Hessian matrix ${\bf {\rm H}}={\bf \nabla}^2 {\Delta_{\rm c}}({\bf x})$ is well defined. Under these conditions, if all minor determinants of ${\bf {\rm H}}$ are non-negative, then ${\Delta_{\rm c}}$ is convex. Explicitly, the entries of ${\bf {\rm H}}$ are given by 
\begin{equation}
{\bf {\rm H}}_{ij}=\frac{\partial^2}{\partial x_i \partial x_j}{\Delta_{\rm c}}({\bf x})=4[\Delta_{\rm c}]_{ij}-\frac{8}{D^2-D}.
\end{equation}
One can easily show that the minor determinant of degree $k$ is 
$${\rm Det}_k=4^k\left(1-\frac{2k}{D^2-D}\right)=4^k\left(1-\frac{k}{M}\right),$$
which is always non-negative, since $D\ge 2$. This finishes the proof. 

Note that for $D=2$ we have ${\rm Det}_k=0$ for all $k=1, \dots, M$, which is consistent with the fact that ${\Delta_{\rm c}}=0$ for all $2\times 2$ states.
For $D \ge 3$ we have ${\rm Det}_k>0$ for  $k=1, \dots, M-1$ and ${\rm Det}_M=0$. Therefore, we have convexity but not strict convexity. This is also expected because ${\Delta_{\rm c}}=0$ for all states for which all coherences have the same absolute value.
\section{States that maximize ${\Delta_{\rm c}}$}
\label{A2}
Here we determine the general states which maximize coherence dispersion. Thanks to the convexity of ${\Delta_{\rm c}}$ one can restrict the search to pure states. If the pure state is of rank 1 in the computational basis, then ${\Delta_{\rm c}}=0$ and, thus, no such a state can maximize the dispersion.
Next, consider rank-2 states, say $|\psi\rangle=\alpha |1\rangle+\beta |2\rangle+0\,(|3\rangle+\dots+|D\rangle)$ and ask which of these states maximize ${\Delta_{\rm c}}$. The average of the absolute values of the coherences is 
\begin{equation}
\label{MediaR2}
 \overline{\rho}_{\rm c}=\frac{2xy}{D^2-D},
\end{equation}
with $x=|\alpha|\ne 0$ and $y=|\beta|\ne 0$. Since we have $D^2-D-2$ vanishing coherences, the dispersion reads
\begin{equation}
{\Delta_{\rm c}}=(D^2-D-2) \overline{\rho}_{\rm c}^2+2(xy- \overline{\rho}_{\rm c})^2.
\end{equation}
Using Eq.(\ref{MediaR2}), we get
\begin{equation}
{\Delta_{\rm c}}=2xy\left(1-\frac{2}{D^2-D}\right),
\end{equation}
which, for $D>2$, is maximized for $x=y=1/\sqrt{2}$. In this simple case we were able to deal with the normalization constraint by inspection. So the rank-2 state that maximizes ${\Delta_{\rm c}}$ in a Hilbert space of arbitrary dimension $D>2$ is of the kind
\begin{equation}
|\psi\rangle=\frac{1}{\sqrt{2}}( |i\rangle+e^{i\varphi} |j\rangle),
\end{equation}
with $i\ne j \in \{1,2,\dots,D\}$ and $\varphi \in [0,2\pi)$.
\subsection{Arbitrary rank}
We now address the general situation of a pure state of fixed rank $r$, with $2<r\le D$. We write 
\begin{equation}
|\psi\rangle=\sum_{i=1}^r\alpha_i|i\rangle,
\end{equation}
and define $x_{i}=|\alpha_i|$, with $i \in\{1,2,\dots, r \}$ and $x_i\ne 0$ (otherwise the rank would differ from $r$). There are $r^2-r$ non-vanishing coherences with $|\rho_{ij}|=x_ix_j$. The average of the absolute values of coherences is
\begin{equation}
\label{MediaR}
 \overline{\rho}_{\rm c}=\frac{1}{D^2-D}\sum_{i\ne j}^{r} x_ix_j,
\end{equation}
and the dispersion becomes
\begin{equation}
{\Delta_{\rm c}}=(D^2-D-r^2+r)\overline{\rho}_{\rm c}^2+\sum_{i\ne j}^{r} \left(x_ix_j-\overline{\rho}_{\rm c}\right)^2.
\end{equation}
The normalization constraint is given by $ \sum_i x_i^2=1$.
We can deal with this condition via a Lagrange multiplier, so that the function to be maximized is 
\begin{equation}
{\cal L}=(D^2-D-r^2+r)\overline{\rho}_{\rm c}^2+\sum_{i\ne j}^{r} \left(x_ix_j-\overline{\rho}_{\rm c}\right)^2-\lambda\left(\sum_{i=1}^{r} x_i^2-1\right).
\end{equation}
By noting that  
\begin{equation}
\frac{\partial \overline{\rho}_{\rm c}}{\partial x_{k}}=\frac{2}{D^2-D}(\varSigma-x_k),\;\; \varSigma\equiv \sum_{i=1}^r x_i,
\end{equation}
after some algebra, the $r$ extremization conditions  $\partial {\cal L}/\partial x_{k}=0$ can be set in the simple form
\begin{equation}
\label{extremal}
1+\overline{\rho}_{\rm c}-\frac{\lambda}{2}=x_k^2+\frac{\overline{\rho}_{\rm c}\varSigma}{x_k}
\end{equation}
for $k=1, 2, \dots, r$. Since the left-hand side of the above equation is a constant, it must be true that
\begin{equation}
x_k^2+\frac{\overline{\rho}_{\rm c}\varSigma}{x_k}=x_j^2+\frac{\overline{\rho}_{\rm c}\varSigma}{x_j} \Rightarrow (x_k-x_j)\left( x_k+x_j-\frac{\overline{\rho}_{\rm c}\varSigma}{x_jx_k}\right)=0
\end{equation}
for all pairs $(k,j)$. Therefore, either $x_k=x_j$, or 
\begin{equation}
x_k+x_j-\frac{\overline{\rho}_{\rm c}\varSigma}{x_jx_k}=0 \Rightarrow  x_jx_k(x_k+x_j)=\overline{\rho}_{\rm c}\varSigma.
\end{equation}
However, since $\overline{\rho}_{\rm c}\varSigma$ is a constant,  this should imply
\begin{equation}
 x_jx_k(x_k+x_j)= x_nx_k(x_k+x_n) \Rightarrow x_j^2-x_n^2=x_k(x_n-x_j)  \Rightarrow x_j+x_n=-x_k,
\end{equation}
which is impossible, since $x_i>0$, leading to a contradiction.

Therefore, the only solution is $|\alpha_1|=|\alpha_2|=\dots=|\alpha_r|\equiv x$. Finally, the last condition  $\partial {\cal L}/\partial \lambda=0$ is simply normalization: $rx^2=1$,
or $x=1/\sqrt{r}$. The Lagrange multiplier is given by 
\begin{equation}
\lambda=2\frac{(r-1)}{r}\left(1-\frac{(r^2-r)}{D^2-D}\right). 
\end{equation}
We conclude that, fixed a rank $r$, with $2<r<D$, the pure state that maximizes ${\Delta_{\rm c}}$ is given by
\begin{equation}
|\psi\rangle=\frac{1}{\sqrt{r}}\sum_{j=1}^re^{i\varphi_j}|j\rangle,
\end{equation}
with  $\varphi_i \in [0,2\pi)$. Of course, any shuffling of basis elements in the ket above leads to a state that also maximizes the dispersion. 
\subsection{Finding the optimal rank}
We are left with the simpler problem of finding the rank that globally maximizes the dispersion, given the dimension $D$. For a fixed dimension $D$, the coherence dispersion 
of the non-full rank states derived in the previous section reads
\begin{equation}
\label{delta}
\Delta_{\rm c}(r)=(D^2-D-r^2+r)\overline{\rho}_{\rm c}^2+(r^2-r)\left(\frac{1}{r}-\overline{\rho}_{\rm c} \right)^2=\left(1-\frac{1}{r} \right)\left(1-\frac{r^2-r}{D^2-D} \right),
\end{equation}
where we used $\overline{\rho}_{\rm c}=(r-1)/(D^2-D)$. Note that the above expression vanishes for both, $r=1$ and $r=D$. We see that $0\le \Delta_{\rm c} \le 1$, where the upper limit is not tight for any finite $D$. 
In fact, we will see that the optimal rank is a slowly increasing function of $D$, so that, only when the dimension becomes arbitrarily large, we have $\Delta_{\rm c} \rightarrow 1$, for the maximal state.

By setting $r \rightarrow s$ and treating $s$ as a continuous variable in this intermediate step, one can require $d\Delta_{\rm c}/ds=0$, leading to the simple cubic equation 
\begin{equation}
s^3-s^2=\frac{D^2-D}{2},
\end{equation}
which has a single real root, given by
\begin{equation}
\mathfrak{s}=\frac{1}{3}\left( 1+\frac{\xi}{2}+\frac{2}{\xi}\right),
\end{equation}
where 
\begin{equation}
\xi=\left(8+54(D^2-D)+6\sqrt{(D^2-D)[15+81(D-1/2)^2]}\right)^{1/3}.
\end{equation}
\begin{figure}
  \includegraphics[height=6.5cm]{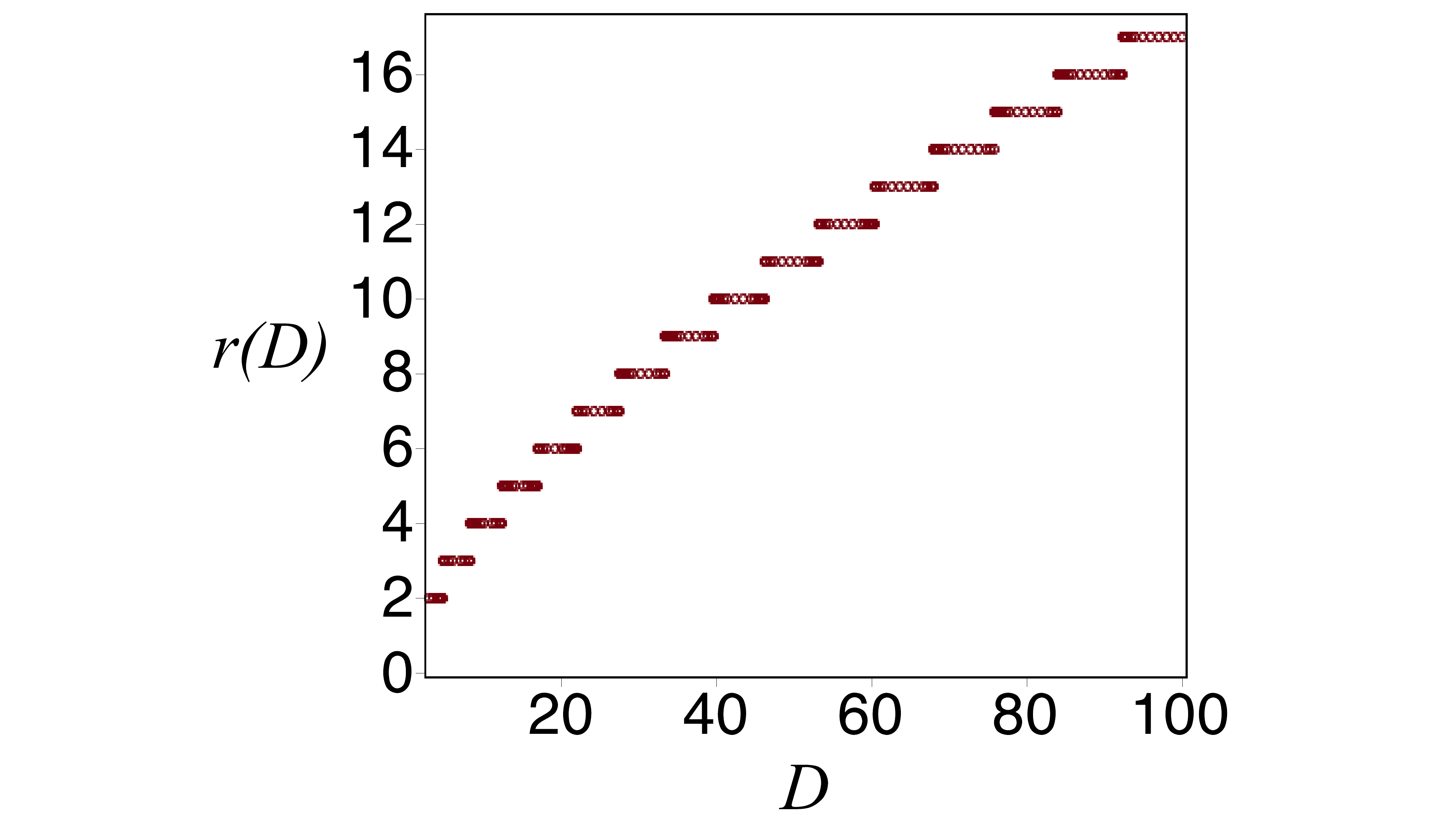}
  \caption{Rank $r(D)$ which maximizes $\Delta_{\rm c}$ as a function of $D$.}
  \label{fig-supp}
\end{figure}

The optimal (integer) rank $r(D)$ is the neighboring integer to $\mathfrak{s}$ which maximizes $\Delta_{\rm c}$. One write the closest smaller integer as $\lfloor \mathfrak{s} \rfloor$ and the closest larger integer as $\lceil \mathfrak{s} \rceil$.
Using equation (\ref{delta}) we get the following difference
\begin{equation}
\Delta_{\rm c}(\lceil \mathfrak{s} \rceil)-\Delta_{\rm c}(\lfloor \mathfrak{s} \rfloor)=\frac{1}{\lceil \mathfrak{s} \rceil\lfloor \mathfrak{s} \rfloor}-\left(\frac{\lceil \mathfrak{s} \rceil+\lfloor \mathfrak{s} \rfloor}{D^2-D} \right)+\frac{2}{D^2-D}\equiv X.
\end{equation}
This quantity oscillates between positive and negative values as $\mathfrak{s}$ varies. Therefore, one can express the optimal rank $r(D)$ as
\begin{equation}
\label{opt}
r(D)= \lfloor \mathfrak{s} \rfloor +\frac{1+{ \rm sign}(X)}{2},
\end{equation}
which is a slowly increasing function of $D$. For larger values of $D$, 
\begin{equation}
r(D)\approx 2^{-1/3}\,D^{2/3}
\end{equation}
is a reasonable approximation to Eq. (\ref{opt}).
We conclude that the states which globally maximize the variance, for a given dimension $D$, can be written as 
\begin{equation}
|\Phi_{\rm M}\rangle=\frac{1}{\sqrt{r(D)}}\sum_{j=1}^{r(D)}e^{i\varphi_j}|j\rangle,
\end{equation}
again, any shuffling of basis elements leads to a state that also maximizes the dispersion. In Fig. \ref{fig-supp} we plot $r(D)$ between $D=3$ and $D=100$. The rank $r(D)$ is a slowly increasing function: for $D=3,4$, we find $r(D)=2$; for $93\le D \le 100$, $r(D)=17$; and for $997 \le D\le 1015$ we have $r(D)=80$. Therefore, e. g., for $D=100$, we have $0 \le \Delta_{\rm c} \le 0.915$.
Whenever other constraints are valid, the form of the maximal states may be quite distinct from the previous equation. For instance, if one requires that the populations
should coincide with the Maxwell-Boltzmann distribution (see the main text for details).

\section{Coherence dispersion of the partially coherent Gibbs state}
\label{A3}
Let us consider the partially coherent Gibbs state $G$, whose entries are
\begin{equation}
\nonumber
G_{jj}=\frac{1}{\cal Z}e^{-\beta E_j}\;\;  \mbox{and} \;\; G_{jk}=\frac{\lambda}{\cal Z}  e^{-\frac{\beta}{2}( E_j+E_k)}\;\;(j\ne k)
\end{equation}
where $\beta=(k_BT)^{-1}$ is the ``reciprocal temperature''  and $\lambda \in [0,1]$ reflects the level of coherence of $G$. 
The purity of $G$ is given by $\Pi=\sum_{j,k}|G_{jk}|^2$, which leads to
\begin{equation}
\Pi=\sum_{j,k}G_{jk}^2=\frac{\lambda^2}{{\cal Z}^2(\beta)}\sum_{i,j}e^{-\beta(E_i+E_j)}+\frac{(1-\lambda^2)}{{\cal Z}^2(\beta)}\sum_ie^{-2\beta E_i}=\lambda^2+(1-\lambda^2)\frac{{\cal Z}(2\beta)}{{\cal Z}^2(\beta)},
\end{equation}
where we used $|G_{jk}|=G_{jk}$ and ${\cal Z}(x \beta)=\sum_je^{- x \beta E_j}$ is the system's partition function at a temperature $T'=T/x$.

Analogously
\begin{equation}
P^2=\frac{{\cal Z}(2\beta)}{{\cal Z}^2(\beta)}-\frac{1}{d} \;\; \mbox{and} \;\; C_1=\lambda\left(\frac{{\cal Z}^2(\beta/2)}{{\cal Z}(\beta)}-1\right).
\end{equation}
Therefore, the single-copy coherence dispersion reads
\begin{equation}
{\Delta_{\rm c}}(G(\beta))=\lambda^2\left\{\left(1-\frac{{\cal Z}(2\beta)}{{\cal Z}^2(\beta)}\right)-\frac{1}{d^2-d}\left(1-\frac{{\cal Z}^2(\beta/2)}{{\cal Z}(\beta)}\right)^2\right\}.
\end{equation}
Now we note that
\begin{equation}
\frac{{\cal Z}(2\beta)}{{\cal Z}^2(\beta)}=\sum_j \frac{e^{-\beta E_j}}{\cal Z(\beta)}\cdot  \frac{e^{-\beta E_j}}{\cal Z(\beta)}.
\end{equation}
But $p_j= \frac{e^{-\beta E_j}}{\cal Z(\beta)}$ is the Maxwell-Boltzmann probability distribution. Therefore, 
\begin{equation}
\frac{{\cal Z}(2\beta)}{{\cal Z}^2(\beta)}=\sum_j p_j^2=\Pi_{\rm eq}(\beta),
\end{equation}
the purity of the thermal equilibrium state at reciprocal temperature $\beta$. Using the same reasoning we obtain
\begin{equation}
\left(\frac{{\cal Z}(\beta)}{{\cal Z}^2(\beta/2)}\right)^{-1}=\frac{1}{\Pi_{\rm eq}(\beta/2)}.
\end{equation}
Therefore
\begin{eqnarray}
\nonumber
{\Delta_{\rm c}}(G(\beta))=\lambda^2\left[\left(1-\Pi_{\rm eq}(\beta)\right)-\frac{1}{d^2-d}\left( 1-1/\Pi_{\rm eq}(\beta/2) \right)^2\right] \\
\Rightarrow {\Delta_{\rm c}}(G(T))=\lambda^2\left[\left(1-\Pi_{\rm eq}(T)\right)-\frac{1}{d^2-d}\left( 1-1/\Pi_{\rm eq}(2T) \right)^2\right] .
\end{eqnarray}
We now assume that the spectrum is that of a $d$-level system, given by $E_j=j\epsilon$. With this we get
\begin{eqnarray}
{\cal Z}(\beta)=e^{-\frac{\beta}{2}(d+1)\epsilon}\,\frac{\sinh(\beta d \epsilon /2)}{\sinh(\beta \epsilon/2)}.
\end{eqnarray}
Therefore,
\begin{eqnarray}
\Pi_{\rm eq}(\beta)&=&\frac{\sinh(\beta d \epsilon) \sinh^2(\beta \epsilon /2)}{\sinh(\beta \epsilon) \sinh^2(\beta d \epsilon /2)}=\tanh( \beta \epsilon/2) \coth(\beta d \epsilon/2)= \tanh( 2/\tau) \coth(2d/\tau),
\end{eqnarray}
where we defined the dimensionless temperature $\tau\equiv 4k_BT/\epsilon$. For a single copy, we obtain
\begin{equation}
{\Delta_{\rm c}}(G(\tau))=\lambda^2\left(1-\tanh( 2/\tau) \coth(2d/\tau)\right)-\frac{\lambda^2}{d^2-d}\left(1-\coth( 1/\tau) \tanh( d/\tau) \right)^2.
\end{equation}
Note that $\lambda^2$ is a global scale factor and, for $d=2$, the previous expression vanishes identically, as it should.

Finally, the multi-copy coherence dispersion reads
\begin{equation}
{\Delta_{\rm c}}(G^{\otimes n}(\beta))= \left(\lambda^2+(1-\lambda^2)\Pi_{\rm eq}(\beta)\right)^{n}-\Pi_{\rm eq}(\beta)^n-\frac{1}{d^{2n}-d^n}\left( \left( 1-\lambda+\lambda /\Pi_{\rm eq}(\beta/2) \right)^n-1
\right)^2,
\end{equation}
which, for $n$ $d$-level systems, in terms of $\tau$, becomes 
\begin{eqnarray}
\nonumber
{\Delta_{\rm c}}(G^{\otimes n}(\tau))&=& \left(\lambda^2+(1-\lambda^2) \tanh( 2/\tau) \coth(2d/\tau)\right)^{n}-\left( \tanh( 2/\tau) \coth(2d/\tau)\right)^n\\
&-&\frac{1}{d^{2n}-d^n}\left(\left(1-\lambda+\lambda\coth( 1/\tau) \tanh( d/\tau)\right)^n-1\right)^2.
\end{eqnarray}
Note that in this expression $\lambda^2$ is no longer a simple multiplicative factor. However, since this is the case for a single copy, the maximum of the multi-copy expression above inherits a quite weak dependence on $\lambda$.

\end{document}